\documentclass[12pt]{article}
\usepackage{amsmath,amssymb,amsthm,amsxtra,overpic,bbm,bm,epsfig,ulem}
\usepackage{color}
\usepackage{cite}

\usepackage{hyperref}
\usepackage{url}

\def\thefootnote{\fnsymbol{footnote}}

\begin{document}

\vspace{0.2cm}

\begin{center}
{\large\bf First determination of the Jarlskog invariant of CP violation \\
from the moduli of the CKM matrix elements}
\end{center}

\vspace{0.2cm}

\begin{center}
{\bf Shu Luo}$^1$
\footnote{E-mail: luoshu@xmu.edu.cn}
and
{\bf Zhi-zhong Xing$^{2,3}$}
\footnote{E-mail: xingzz@ihep.ac.cn}
\\
{\small $^{1}$Department of Astronomy, Xiamen University, Fujian 361005, China} \\
{\small $^{2}$Institute of High Energy Physics and School of Physical Sciences, \\
University of Chinese Academy of Sciences, Beijing 100049, China \\
$^{3}$Center of High Energy Physics, Peking University, Beijing 100871, China}
\end{center}

\vspace{2cm}
\begin{abstract}
We find that the precision and accuracy of current experimental data on the
moduli of nine Cabibbo-Kobayashi-Maskawa (CKM) quark flavor mixing matrix
elements allow us to numerically determine the {\it correct} size of the Jarlskog
invariant of CP violation from four of them in eight different ways {\it for the
first time} without making any special assumptions. This observation implies a 
remarkable self-consistency of the correlation between CP-conserving and CP-violating 
quantities of the CKM matrix as guaranteed by its unitarity.
\end{abstract}

\newpage

\def\thefootnote{\arabic{footnote}}
\setcounter{footnote}{0}
\setcounter{figure}{0}

\section{Introduction}

In the standard model (SM) of particle physics, the phenomena of quark
flavor mixing and weak CP violation are elegantly described by a nontrivial
$3\times 3$ unitary matrix appearing in the flavor-changing charged-current
interactions, the well-known Cabibbo-Kobayashi-Maskawa (CKM) matrix
$V$~\cite{Cabibbo:1963yz,Kobayashi:1973fv}. The unitarity
of $V$, which can be expressed as a combination of the normalization
and orthogonality conditions (for $\alpha, \beta = u, c, t$ and
$i, j = d, s, b$)
\begin{eqnarray}
\sum_i V^{}_{\alpha i} V^*_{\beta i} = \delta^{}_{\alpha \beta} \; ,
\quad\quad
\sum_\alpha V^{}_{\alpha i} V^*_{\alpha j} = \delta^{}_{ij} \; ,
\label{1}
\end{eqnarray}
is the only but powerful constraint imposed by the SM itself. In
particular, this constraint leads us to a unique rephasing-invariant
measure of CP violation in the quark sector --- the so-called Jarlskog
invariant $\cal J$~\cite{Jarlskog:1985ht,Wu:1985ea}
defined through
\begin{eqnarray}
{\rm Im}\left(V^{}_{\alpha i} V^{}_{\beta j} V^*_{\alpha j} V^*_{\beta i}
\right) = {\cal J} \sum_\gamma \epsilon^{}_{\alpha\beta\gamma} \sum_k
\epsilon^{}_{ijk} \; ,
\label{2}
\end{eqnarray}
where $\epsilon^{}_{\alpha\beta\gamma}$ and $\epsilon^{}_{ijk}$ denote
the three-dimensional Levi-Civita symbols with the Greek and Latin subscripts
running respectively over $\left(u, c, t\right)$ and $\left(d, s, b\right)$.
As the moduli of nine CKM matrix elements are also rephasing-invariant,
Eqs.~(\ref{1}) and (\ref{2}) give rise to a rather striking correlation
between the CP-violating and CP-conserving invariants of
$V$~\cite{Sasaki:1986jv,Hamzaoui:1988sh}:
\begin{eqnarray}
{\cal J}^2 =
|V^{}_{\alpha i}|^2 |V^{}_{\beta j}|^2 |V^{}_{\alpha j}|^2 |V^{}_{\beta i}|^2
- \frac{1}{4} \left(|V^{}_{\alpha i}|^2 |V^{}_{\beta j}|^2 +
|V^{}_{\alpha j}|^2 |V^{}_{\beta i}|^2 - |V^{}_{\gamma k}|^2\right)^2 \; ,
\label{3}
\end{eqnarray}
where $\alpha \neq \beta \neq \gamma$ and $i \neq j \neq k$
are certainly required. One may in principle
use this algebraic relation to calculate the size of $\cal J$ from any
four independent moduli of the CKM matrix elements, and then compare the
result with the value of $\cal J$ extracted from the CP-violating asymmetries
in some hadronic decay modes (e.g., an asymmetry between the rates
of $B^0_d$ and $\bar{B}^0_d \to J/\psi + K^{}_{\rm S}$
decays~\cite{ParticleDataGroup:2022pth}). Such a test of the validity of
Eq.~(\ref{3}) with the relevant experimental data makes sense because it
offers another viable way to cross check the unitarity of $V$.

But Eq.~(\ref{3}) has never been successfully confronted with the available 
experimental data in the past decades
\footnote{A preliminary attempt was made to calculate $\cal J$ from the 
inputs of $|V^{}_{us}|$, $|V^{}_{cd}|$, $|V^{}_{ub}|$ and $|V^{}_{cb}|$ in 
Ref.~\cite{Botella:2005fc}, with a conclusion that ``it is not feasible 
{\it in practice}". The authors assumed the Gaussian probability density 
distributions around the central values of the moduli of these four CKM 
matrix elements, and adopted a toy Monte Carlo method to compute the 
probability density distribution of ${\cal J}^2$. They found that only $7.9\%$
of the generated points could assure the positivity of ${\cal J}^2$,
unfortunately.}.
The main reason is simply that the
expression of ${\cal J}^2$ is a difference between two positive terms
consisting of a number of moduli of the CKM matrix elements.
So the positivity and smallness of ${\cal J}^2$ implies that its first
term must be slightly larger than its second term, and a significant
cancellation between these two terms is in general unavoidable. In this
case the input values of all the CP-conserving quantities on the right-hand
side of Eq.~(\ref{3}) must be as precise as possible and maximally compatible
with the unitarity conditions of $V$, otherwise the output value of ${\cal J}^2$
would be either negative or in conflict with the result of $\cal J$ determined
from CP violation in $B$ and $K$ decays. Then the 
question arises: can today's experimental measurements of the CKM matrix 
elements allow us to reliably calculate the size of $\cal J$ from Eq.~(\ref{3})?

The answer is affirmative, but it depends highly on which four independent
$|V^{}_{\alpha i}|$ (for $\alpha = u, c, t$ and $i = d, s, b$) are taken
into account. In this article we are going to show that, {\it for
the first time}, the correct size of the Jarlskog invariant of CP violation
can be numerically calculated from the moduli of the CKM matrix elements
with the available experimental data and without making any special assumptions. 
We find that there are eight different ways to do so for the time being, 
and they all include the moduli of the two smallest CKM matrix elements 
$|V^{}_{ub}|$ and $|V^{}_{td}|$.

\section{How good are the data of $|V^{}_{\alpha i}|$ to fit unitarity?}

To clearly see the fine difference between any two of the CKM matrix elements
with a comparable magnitude, let us adopt the Wolfenstein-like expansion
of $V$ as follows~\cite{Wolfenstein:1983yz,Buras:1994ec}:
\begin{eqnarray}
&& V^{}_{ud} = 1 - \frac{1}{2}\lambda^2 - \frac{1}{8}\lambda^4 + {\cal O}(\lambda^6) \; ,
\nonumber \\
&& V^{}_{us} = \lambda + {\cal O}(\lambda^7) \; ,
\nonumber \\
&& V^{}_{ub} = A\lambda^3 \left(\rho - {\rm i}\eta\right) \; ;
\nonumber \\
&& V^{}_{cd} = -\lambda + \frac{1}{2} A^2 \lambda^5 \left[1 - 2\left(\rho + {\rm i}
\eta\right)\right] + {\cal O}(\lambda^7) \; ,
\nonumber \\
&& V^{}_{cs} = 1 - \frac{1}{2} \lambda^2 - \frac{1}{8} \lambda^4 \left(1 + 4 A^2\right)
+ {\cal O}(\lambda^6) \; ,
\nonumber \\
&& V^{}_{cb} = A \lambda^2 + {\cal O}(\lambda^8) \; ;
\nonumber \\
&& V^{}_{td} = A\lambda^3 \left(1 -\rho - {\rm i}\eta\right) + \frac{1}{2}
A\lambda^5 \left(\rho + {\rm i}\eta\right) + {\cal O}(\lambda^7) \; , \hspace{0.5cm}
\nonumber \\
&& V^{}_{ts} = -A\lambda^2 + \frac{1}{2} A\lambda^4 \left[1 - 2\left(\rho +
{\rm i}\eta\right)\right] + {\cal O}(\lambda^6) \; ,
\nonumber \\
&& V^{}_{tb} = 1 - \frac{1}{2} A^2\lambda^4 + {\cal O}(\lambda^6) \; ,
\label{4}
\end{eqnarray}
where $V^{}_{ub} = A \lambda^3 \left(\rho - {\rm i}\eta\right)$ is exact
by definition, $\lambda$ denotes the small expansion parameter, and the
unitarity of $V$ is valid at the level of ${\cal O}(\lambda^6)$. Current
experimental data~\cite{ParticleDataGroup:2022pth} lead us to
$\lambda \simeq 0.225$, $A \simeq 0.825$, $\rho \simeq 0.163$ and
$\eta \simeq 0.357$ in the neglect of their corresponding error bars.
Then we can easily arrive at a remarkable ordering for the
nine elements of $V$, as first observed in Ref.~\cite{Xing:1996it}:
\begin{eqnarray}
|V^{}_{tb}| > |V^{}_{ud}| > |V^{}_{cs}|
\hspace{-0.2cm} & \gg & \hspace{-0.2cm}
|V^{}_{us}| > |V^{}_{cd}|
\nonumber \\
\hspace{-0.2cm} & \gg & \hspace{-0.2cm}
|V^{}_{cb}| > |V^{}_{ts}|
\nonumber \\
\hspace{-0.2cm} & \gg & \hspace{-0.2cm}
|V^{}_{td}| > |V^{}_{ub}| \; . \hspace{0.5cm}
\label{5}
\end{eqnarray}
In comparison, the present experimental values of nine moduli of $V$
are~\cite{ParticleDataGroup:2022pth}
\begin{eqnarray}
\left ( \begin{matrix} |V^{}_{ud}| & |V^{}_{us}| & |V^{}_{ub}| \cr
|V^{}_{cd}| & |V^{}_{cs}| & |V^{}_{cb}| \cr
|V^{}_{td}| & |V^{}_{ts}| & |V^{}_{tb}| \cr \end{matrix} \right )
= \left ( \begin{matrix} 0.97373 \pm 0.00031 & 0.2243 \pm 0.0008
& ( 3.82 \pm 0.20 ) \times 10^{-3}_{} \cr 0.221 \pm 0.004
& 0.975 \pm 0.006 & ( 40.8 \pm 1.4 ) \times 10^{-3}_{} \cr
( 8.6 \pm 0.2 ) \times 10^{-3}_{} & ( 41.5 \pm 0.9 ) \times 10^{-3}_{}
& 1.014 \pm 0.029 \cr \end{matrix} \right ) \; .
\label{6}
\end{eqnarray}
Some immediate comments are in order.
\begin{itemize}
\item     The experimental values of $|V^{}_{ub}|$ and $|V^{}_{td}|$ confirm
that they are the two smallest moduli of the CKM matrix elements and in the
correct ordering. It will therefore be safe to choose these two
moduli to calculate the magnitude of the Jarlskog invariant $\cal J$ with
the help of Eq.~(\ref{3}), as their impacts on the normalization conditions
of $V$ are negligible in most cases.

\item     The central values of $|V^{}_{cb}|$ and $|V^{}_{ts}|$ imply that
they seem to be in a wrong ordering, in conflict with the expectation shown
in Eq.~(\ref{5}) as required by the unitarity of $V$. So a further improvement
of the precision and accuracy associated with the individual measurements of
$|V^{}_{cb}|$ and $|V^{}_{ts}|$ is no doubt necessary.

\item     The fact that $|V^{}_{us}|$ should be slightly larger than
$|V^{}_{cd}|$ has essentially been established from today's data, as one can
see from Eq.~(\ref{6}). A precision measurement of $|V^{}_{cd}|$ in the
near future may more convincingly strengthen this observation.

\item     The central values of $|V^{}_{ud}|$ and $|V^{}_{cs}|$
imply that they seem to be in an ordering inconsistent with the expectation
from Eq.~(\ref{5}). The reason is simply that there remain some quite
large uncertainties associated with the determination of $|V^{}_{cs}|$. As
both $|V^{}_{ud}|$ and $|V^{}_{cs}|$ are close to one, their errors may
easily invalidate the normalization conditions of $V$ in some cases.

\item     The value of $|V^{}_{tb}|$ involves the largest uncertainty and
is apparently incompatible with the unitarity requirement of $V$, although
it looks like the largest moduli as expected among the nine moduli of the CKM
matrix elements. So one should better avoid using the present experimental 
result of $|V^{}_{tb}|$ to calculate ${\cal J}^2$ via Eq.~(\ref{3}).

\end{itemize}
In short, $|V^{}_{ub}|$ and $|V^{}_{td}|$ should be taken into account
when combining Eq.~(\ref{3}) and Eq.~(\ref{6}) to calculate $\cal J$.
Whether such a calculation can successfully lead us to a meaningful result
of $\cal J$ depends on whether the input values of the other two independent
moduli of the CKM matrix elements are accurate enough and maximally consistent
with the unitarity conditions. Let us make things clear by checking all the
possibilities along this line of thought.

\section{Calculations of $|\cal J|$ from the moduli of $V^{}_{\alpha i}$}

It is straightforward to figure out that there are totally
$\displaystyle C^{4}_{9} = 9! / ( 4! \; 5! ) = 126$ possibilities to randomly
choose any four of the nine CKM matrix elements, but 45 of them should be
abandoned since the chosen four matrix elements are not completely independent.
To be more specific, we find that the possibilities in the following two
categories ought to be eliminated.
\begin{itemize}
\item     If three of the four chosen CKM matrix elements lie in the same
row or column of $V$, they must satisfy the corresponding normalization
condition and thus are not fully independent. There are totally $6 \times 6$
possibilities belonging to this category, where the first ``6" means a sum
of three possible rows and three possible columns, and the second ``6"
indicates that the fourth CKM matrix element may be any of the other six
CKM matrix elements which is located in a different row or column.

\item     If two of the four chosen CKM matrix elements lie in a row of $V$
and the other two lie in a column of $V$ except the possibilities that
three of them are located in the same row or column, then they must not be
fully independent. For example, $V^{}_{us}$ and $V^{}_{ub}$ in the first
row are related to $V^{}_{cd}$ and $V^{}_{td}$ in the first column via
$|V^{}_{us}|^2 + |V^{}_{ub}|^2 = |V^{}_{cd}|^2 + |V^{}_{td}|^2$.
There are totally 9 possibilities of this category.
\end{itemize}
As a result, we are left with $126 - 36 - 9 = 81$ different ways of choosing
a set of four independent CKM matrix elements. We find that these 81
possibilities can be categorized into the following three different groups.
\begin{itemize}
\item     The four chosen CKM matrix elements are independent and located in
{\it two} rows and {\it two} columns of $V$, such as the patterns
\begin{eqnarray}
\left ( \begin{matrix} \times & \times & \hspace{0.33cm} \cr \times & \times & ~ \cr
~ & ~ & ~ \cr \end{matrix} \right ) \; ,
\quad\quad
\left ( \begin{matrix} \times & \hspace{0.33cm} & \times \cr
\times & ~ & \times \cr ~ & ~ & ~ \cr \end{matrix} \right ) \; ,
\quad\quad
\left ( \begin{matrix} \times & \times & \hspace{0.33cm} \cr
~ & ~ & ~ \cr \times & \times & ~ \cr \end{matrix} \right ) \; .
\label{7}
\end{eqnarray}
There are totally 9 different patterns of this category.

\item     The four chosen CKM matrix elements are independent and located in
{\it two} rows and {\it three} columns (or {\it two} columns and {\it three} rows)
of $V$, such as the patterns
\begin{eqnarray}
\left ( \begin{matrix} \times & \times & \hspace{0.3cm} \cr
\times & ~ & \times \cr ~ & ~ & ~ \cr \end{matrix} \right ) \; ,
\quad\quad
\left ( \begin{matrix} \times & \hspace{0.3cm} & \times \cr
~ & ~ & ~ \cr \times & \times & ~ \cr \end{matrix} \right ) \; ,
\quad\quad
\left ( \begin{matrix} \times & \times & \hspace{0.3cm} \cr
\times & ~ & ~ \cr ~ & \times & ~ \cr \end{matrix} \right ) \; .
\label{8}
\end{eqnarray}
There are totally 36 different patterns of this category.

\item     The four chosen CKM matrix elements are independent and located in
{\it three} rows and {\it three} columns of $V$, such as the patterns
\begin{eqnarray}
\left ( \begin{matrix} \times & \times & ~ \cr
\times & ~ & ~ \cr ~ & ~ & \times \cr \end{matrix} \right ) \; ,
\quad\quad
\left ( \begin{matrix} \times & ~ & \times \cr
~ & \times & ~ \cr \times & ~ & ~ \cr \end{matrix} \right ) \; ,
\quad\quad
\left ( \begin{matrix} \times & ~ & \times \cr
\times & ~ & ~ \cr
~ & \times & ~ \cr \end{matrix} \right ) \; .
\label{9}
\end{eqnarray}
There are totally 36 different patterns of this category.
\end{itemize}
As pointed out in section 2, the values of some of the moduli of the nine
CKM matrix elements involve quite large uncertainties and may not
respect the unitarity conditions to a good degree of accuracy when they
are input to calculate ${\cal J}^2$ from Eq.~(\ref{3}). In this case the
output of ${\cal J}^2$ is likely to be either negative or too far away from the
global fit result ${\cal J} = \left(3.08^{+0.15}_{-0.13}\right) \times 10^{-5}$
advocated by the Particle Data Group~\cite{ParticleDataGroup:2022pth}.

We proceed to do a careful numerical analysis of all the aforementioned 
81 possibilities of choosing the four independent CKM matrix elements and 
calculating ${\cal J}^2$ from Eq.~(\ref{3}) by adopting a strategy as
follows. Given the very fact that the errors of $|V^{}_{\alpha i}|$ (for
$\alpha = u, c, t$ and $i = d, s, b$) listed in Eq.~(\ref{6}) involve 
some theoretical uncertainties which do not really obey the Gaussian 
probability density distributions~\cite{ParticleDataGroup:2022pth}, 
we simply make a random scan within the given error bar for each of the nine
$|V^{}_{\alpha i}|$ instead of assuming any particular statistical  
distributions regarding the uncertainties of $|V^{}_{\alpha i}|$. This
conservative strategy may largely assure that the correct output of ${\cal J}^2$
is not fragile in the sense that it is essentially stable even when a particular 
probability density distribution around the central value of $|V^{}_{\alpha i}|$ 
is assumed (but the reverse may not be true, as we have checked). 
After some lengthy calculations, we find that current data on $|V^{}_{\alpha i}|$ 
only allow the following eight choices to be viable.
\begin{itemize}
\item     The four independent CKM matrix elements are $V^{}_{ud}$, $V^{}_{ub}$,
$V^{}_{td}$ and $V^{}_{cb}$ or $V^{}_{ts}$:
\begin{eqnarray}
\left ( \begin{matrix} V^{}_{ud} & \hspace{0.4cm} & V^{}_{ub} \cr
~ & ~ & V^{}_{cb} \cr V^{}_{td} & ~ & ~ \cr \end{matrix} \right ) \; ,
\quad\quad
\left ( \begin{matrix} V^{}_{ud} & \hspace{0.4cm} & V^{}_{ub} \cr
~ & ~ & ~ \cr V^{}_{td} & V^{}_{ts} & ~ \cr \end{matrix} \right ) \; ,
\label{10}
\end{eqnarray}
from which the results ${\cal J} = \left(3.20^{+0.25}_{-0.28}\right) \times 10^{-5}$
and ${\cal J} = \left(3.19^{+0.25}_{-0.32}\right) \times 10^{-5}$ can
be respectively obtained
\footnote{We have abandoned the respective solutions
${\cal J} = -\left(3.20^{+0.25}_{-0.28}\right) \times 10^{-5}$ and 
${\cal J} = -\left(3.19^{+0.25}_{-0.32}\right) \times 10^{-5}$ that are mathematically
allowed by Eq.~(\ref{3}), simply because ${\cal J} > 0$ has been experimentally 
established on solid ground~\cite{ParticleDataGroup:2022pth}.}.
Here the central value of ${\cal J}$ is achieved from the central values 
of the four input moduli, and its upper and lower bounds correspond to 
its maximal and minimal values extracted from our random scans within the given 
error bars of the relevant moduli.  

\item     The four independent CKM matrix elements are $V^{}_{us}$, $V^{}_{ub}$,
$V^{}_{td}$ and $V^{}_{cb}$ or $V^{}_{ts}$:
\begin{eqnarray}
\left ( \begin{matrix}  & V^{}_{us} & V^{}_{ub} \cr
~ & ~ & V^{}_{cb} \cr V^{}_{td} & ~ & ~ \cr \end{matrix} \right ) \; ,
\quad\quad
\left ( \begin{matrix}  & V^{}_{us} & V^{}_{ub} \cr
~ & ~ & ~ \cr V^{}_{td} & V^{}_{ts} & ~ \cr \end{matrix} \right ) \; ,
\label{11}
\end{eqnarray}
from which the numerical results ${\cal J} = \left(3.19^{+0.25}_{-0.25}\right) 
\times 10^{-5}$ and ${\cal J} = \left(3.20^{+0.25}_{-0.29}\right) \times 10^{-5}$ can
be respectively achieved.

\item     The four independent CKM matrix elements are $V^{}_{cd}$, $V^{}_{ub}$,
$V^{}_{td}$ and $V^{}_{cb}$ or $V^{}_{ts}$:
\begin{eqnarray}
\left ( \begin{matrix}  & \hspace{0.4cm} & V^{}_{ub} \cr
V^{}_{cd} & ~ & V^{}_{cb} \cr V^{}_{td} & ~ & ~ \cr \end{matrix} \right ) \; ,
\quad\quad
\left ( \begin{matrix}  & \hspace{0.4cm} & V^{}_{ub} \cr
V^{}_{cd} & ~ & ~ \cr V^{}_{td} & V^{}_{ts} & ~ \cr \end{matrix} \right ) \; ,
\label{12}
\end{eqnarray}
from which ${\cal J} = \left(3.19^{+0.26}_{-0.26}\right) \times 10^{-5}$
and ${\cal J} = \left(3.20^{+0.24}_{-0.29}\right) \times 10^{-5}$ can
be respectively obtained.

\item     The four independent CKM matrix elements are $V^{}_{cs}$, $V^{}_{ub}$,
$V^{}_{td}$ and $V^{}_{cb}$ or $V^{}_{ts}$:
\begin{eqnarray}
\left ( \begin{matrix}  &  & V^{}_{ub} \cr
~ & V^{}_{cs} & V^{}_{cb} \cr V^{}_{td} & ~ & ~ \cr \end{matrix} \right ) \; ,
\quad\quad
\left ( \begin{matrix}  &  & V^{}_{ub} \cr
~ & V^{}_{cs} & ~ \cr V^{}_{td} & V^{}_{ts} & ~ \cr \end{matrix} \right ) \; ,
\label{13}
\end{eqnarray}
from which ${\cal J} = \left(3.18^{+0.27}_{-0.55}\right) \times 10^{-5}$
and ${\cal J} = \left(3.20^{+0.25}_{-0.58}\right) \times 10^{-5}$ can
be respectively achieved.
\end{itemize}
The most salient and common feature of these eight patterns is that they all
include the two smallest CKM matrix elements $V^{}_{ub}$ and $V^{}_{td}$ of
${\cal O}(\lambda^3)$, besides one of the CKM matrix elements of
${\cal O}(\lambda^2)$ (i.e., $V^{}_{cb}$ or $V^{}_{ts}$). In any of the above
eight choices, the fourth CKM matrix element can be either $V^{}_{ud}$ (or
$V^{}_{cs}$) of ${\cal O}(1)$ or $V^{}_{us}$ (or $V^{}_{cd}$) of ${\cal O}(\lambda)$.
As expected, the possibilities associated with the largest CKM matrix element
$V^{}_{tb}$ have been excluded from our calculations simply because
the present value of $|V^{}_{tb}|$ is most ``unitarity-unfriendly".
One may also see that the outputs of $\cal J$ involve a bit larger error
bars in Eq.~(\ref{13}) as compared with those in Eqs.~(\ref{10})---(\ref{12}),
since the input value of $|V^{}_{cs}|$ remains ``unitarity-unsatisfactory".

At this point one may expect that the allowed range of ${\cal J}$ in each of
the above eight cases should more or less be narrowed, if the additional
constraints $|V^{}_{td} / V^{}_{ts}| = 0.207 \pm 0.004$ and 
$|V^{}_{ub} / V^{}_{cb}| = 0.084 \pm 0.007$~\cite{ParticleDataGroup:2022pth} 
are taken into account together with Eq.~(\ref{6}). We confirm that this 
expectation is true, and obtain
${\cal J} = \left(3.20^{+0.25}_{-0.28}\right) \times 10^{-5}$
and ${\cal J} = \left(3.19^{+0.25}_{-0.27}\right) \times 10^{-5}$ for 
the two patterns in Eq.~(\ref{10}); 
${\cal J} = \left(3.19^{+0.25}_{-0.25}\right) \times 10^{-5}$
and ${\cal J} = \left(3.20^{+0.25}_{-0.26}\right) \times 10^{-5}$ 
for the two patterns in Eq.~(\ref{11}); 
${\cal J} = \left(3.19^{+0.26}_{-0.25}\right) \times 10^{-5}$
and ${\cal J} = \left(3.20^{+0.24}_{-0.26}\right) \times 10^{-5}$ 
for the two patterns in Eq.~(\ref{12});
${\cal J} = \left(3.18^{+0.27}_{-0.52}\right) \times 10^{-5}$
and ${\cal J} = \left(3.20^{+0.25}_{-0.47}\right) \times 10^{-5}$ 
for the two patterns in Eq.~(\ref{13}). In particular,
the lower bound of ${\cal J}$ is more sensitive to the constraint 
$|V^{}_{td} / V^{}_{ts}|$. 

\section{Summary}

The usefulness of the Jarlskog invariant $\cal J$ as a rephasing-independent
measure of weak CP violation has been well recognized in both the
quark sector and the lepton sector
\footnote{This will be true if the $3\times 3$ Pontecorvo-Maki-Nakagawa-Sakata (PMNS)
lepton flavor mixing matrix~\cite{Pontecorvo:1957cp,Maki:1962mu,Pontecorvo:1967fh}
is assumed to be exactly unitary. The early discussions about leptonic CP violation
in a $3\times 3$ unitary flavor mixing matrix can be found in
Ref.~\cite{Kobayashi:1977ss} and especially in Ref.~\cite{Cabibbo:1977nk}.}.
It is also known that $\cal J$ can be expressed in terms of any four independent
moduli of the nine quark or lepton flavor mixing matrix elements, as guaranteed
by the unitarity conditions. This kind of correlation between the CP-violating and
CP-conserving quantities should be experimentally tested, as it can provide a
novel way to cross check the unitarity of the CKM or PMNS matrix.

We have shown that, {\it for the first time}, the correct size of the Jarlskog
invariant of CP violation can be numerically calculated from the moduli of the
CKM matrix elements with the help of the currently available experimental data
and without making any special assumptions.
But we find that there are only eight different ways to do so, as limited by
the precision and accuracy of the relevant experimental values of
$|V^{}_{\alpha i}|$ (for $\alpha = u, c, t$ and $i = d, s, b$).
This encouraging observation implies that the unitarity of the CKM matrix
deserves a further and more reliable test in the upcoming precision measurement
era of flavor physics characterized by the High-Luminosity Large Hadron
Collider. The same expectation makes sense for testing the unitarity
of the $3 \times 3$ PMNS matrix and constraining possible extra species of 
massive neutrinos in the precision measurement era of neutrino physics
\footnote{The possibility of determining the Jarlskog invariant of leptonic
CP violation from the three CP-conserving quantities in $\nu^{}_\mu \to \nu^{}_e$
and $\overline{\nu}^{}_\mu \to \overline{\nu}^{}_e$ oscillations has
recently been discussed by us~\cite{Luo:2023xmv}.}.

\vspace{0.3cm}

{\it This work is supported in part by the National Natural Science Foundation
of China under grant No. 11775183 (S.L.) and grant Nos. 12075254 and
11835013 (Z.Z.X.).}

\end{document}